\documentclass[final,5p,times,twocolumn]{elsarticle}

\usepackage{amsmath}
\usepackage{amssymb}

\biboptions{compress}

\renewcommand{\d}{{\mathrm{d}}}
\newcommand{\e}{{\mathrm{e}}}
\newcommand{\rmax}{{r_\mathrm{max}}}
\newcommand{\rmin}{{r_\mathrm{min}}}
\newcommand{\f}{_\mathrm{f}}
\newcommand{\muf}{\mu}
\renewcommand{\c}{_\mathrm{th}}
\newcommand{\Prob}{P}
\newcommand{\Probd}{p}
\newcommand{\Probdc}{p_\mathrm{th}}
\newcommand{\nc}{_\mathrm{nc}}
\renewcommand{\b}{_\mathrm{b}}
\newcommand{\crit}{^\mathrm{c}}
\newcommand{\sigmacritbundle}{\sigma\crit_\mathrm{bundle}}
\newcommand{\weibullexp}{k}

\begin{document}

\begin{frontmatter}

\title{Breakdown of fiber bundles with stochastic load-redistribution}

\author{J\"org Lehmann}
\ead{Joerg.Lehmann@ch.abb.com}
\author{Jakob Bernasconi}
\address{ABB Switzerland Ltd., Corporate Research, Segelhofstrasse 1K, CH-5405~Baden-D\"attwil, Switzerland}

\begin{abstract}
  We study fracture processes within a stochastic fiber-bundle model
  where it is assumed that after the failure of a fiber, each intact fiber obtains a
  random fraction of the failing load. Within a Markov approximation,
  the breakdown properties of this model can be reduced to the
  solution of an integral equation. As examples we consider two
  different versions of this model that both can interpolate between
  global and local load redistribution. For the strength thresholds
  of the individual fibers, we consider a Weibull distribution and a
  uniform distribution, both truncated below a given initial stress.
  The breakdown behavior of our models is compared with corresponding
  results of other fiber-bundle models.
\end{abstract}

\begin{keyword}
fracture mechanics \sep fiber-bundle model \sep statistical physics \sep branching process 
\end{keyword}

\end{frontmatter}

\section{Introduction}

\label{sec:intro}

Fracture processes in heterogeneous materials are an important
technological problem that has attracted the interest of the
scientific community since a long
time~\cite{Herrmann1990,Chakrabarti1997,Alava2006,Pradhan2008}. Due to the
complex interaction between failures and the subsequent redistribution
of local stresses, the development of adequate statistical models for
fracture propagation is an extremely hard and challenging
undertaking. The probably most important class of approaches to the
study of fracture processes is that of fiber-bundle models
(FBM's)~\cite{Daniels1945,Kloster1997,Bhattacharyya2003,Zhang1996,Wu2000,Kun2000,Raischel2006,Hidalgo2002,Pradhan2005,Curtin1998,Dalton2009}. Despite
their simplicity, FBM's are able to describe the main processes that
can lead to a propagation of fractures and eventually to a complete
breakdown of real heterogeneous materials.

Fiber-bundle models refer to a bundle of $N$ parallel fibers that are clamped
at both ends and stretched by a common force~$F$. The fibers have a
stochastic distribution of individual strength thresholds, and the
different versions of FBM's that have been considered can be
distinguished by their assumptions with respect to the
stress redistribution after the failure of one of the fibers. The
usual experimental setup considered and analyzed in the FBM literature
can be described as follows: The force~$F$ is gradually increased from
zero until the weakest fiber breaks, and the transfer of its stress to
the surviving fibers may then induce an avalanche of subsequent
failures. If the fiber bundle reaches an equilibrium with no further
failures, the force~$F$ is increased again until the next fiber breaks,
and this procedure is repeated up to the complete breakdown of the
entire bundle. The main quantities of interest in connection with this
procedure are the distribution of avalanche sizes and the ultimate
strength of the fiber bundle, defined as the maximum stress~$F/N$ the system
can support before it breaks down completely. An alternative but
equivalent procedure is to apply a finite force~$F$ to the system, so
that immediately all fibers with a strength threshold smaller than $F/N$
fail. The ultimate strength of the fiber bundle is then determined by
the maximum $F/N$ value that does not lead to a failure of the entire system.

The oldest and most well-known FBM is that where the stress of a
failing fiber is distributed equally between the surviving fibers
(global load sharing, GLS)~\cite{Daniels1945}. For this model, the
strength of the fiber bundle as well as the form of the avalanche-size
distribution can be determined
analytically~\cite{Kloster1997,Bhattacharyya2003}.

Local load sharing (LLS) fiber-bundle models, on the other hand, are
much more difficult to analyze~\cite{Kloster1997,Zhang1996,Wu2000}. In these
models, the stress of a failing fiber is only transferred locally,
typically to the surviving nearest neighbors. LLS models have been
studied mainly via Monte-Carlo simulations and analytical results have
only been obtained for one-dimensional models with essentially nearest-neighbor load transfer.

In most studies, the strength thresholds of the individual fibers are
assumed to be distributed according to a Weibull distribution
(typically with a Weibull index $\weibullexp=2$). For simplicity, however,
uniform strength-threshold distributions (sometimes with a finite
lower cutoff) have also been considered~\cite{Kloster1997,Bhattacharyya2003,Raischel2006}.

The two idealized extremes of global load sharing and of load transfer
to nearest neighbors only are not adequate assumptions for most real
systems. Attempts have therefore been made to interpolate between GLS
and LLS behavior~\cite{Kun2000,Raischel2006,Hidalgo2002,Pradhan2005}. Hidalgo et
al.~\cite{Hidalgo2002}, e.g., assume that the stress transfer after a
failure decays as $1 / r^\gamma$, where $r$ is the distance from the
broken fiber, and they study the failure-propagation behavior of such
a system as a function of $\gamma$. A similar model (with $\gamma =
3$) has been studied by Curtin~\cite{Curtin1998}.

In Ref.~\cite{Lehmann2009}, we have introduced a class of failure-propagation
models that can represent, in a stochastic sense, the main
characteristics of realistic load-redistribution mechanisms, but are
still amenable to an analytical treatment. We have applied our
approach to an illustrative prototype example for cascading failure
propagation in large infrastructure networks, e.g., power grids. In
particular, we have analyzed the probability of a complete system
breakdown after an initial failure, and we have found that the model
exhibits interesting critical dependencies on parameters that
characterize the failure tolerance of the individual elements and the
range of load redistribution after a failure.

In this paper, we apply our stochastic approach to the problem of
fracture propagation in fiber bundles and analyze two models that
interpolate between global and local stress redistribution. The first
is a stochastic version of the $r^{-\gamma}$-model of Hidalgo et
al.~\cite{Hidalgo2002}, and the second a fiber-bundle equivalent to
our prototype example of Ref.~\cite{Lehmann2009}.  We consider an
experimental setup where initially all fibers carry the same stress
$\sigma_0$ and where the individual strength thresholds $\sigma\c$ are
randomly distributed according to a probability density
$\Probdc(\sigma\c)$ that is zero for $\sigma\c < \sigma_0$. We then
examine the consequences of the breaking of a single fiber and
concentrate on the calculation of the following quantities:
\begin{enumerate}[(i)]
\item the \textit{no-cascade probability}~$\Prob\nc(\sigma_0)$, i.e.,
  the probability that an initial failure does not induce any further
  failures;
\item the \textit{breakdown probability}~$\Prob\b(\sigma_0)$, i.e.,
  the probability that an initial failure leads to a breakdown of the
  entire fiber bundle; and
\item the \textit{critical stress}~$\sigma_0\crit$, defined as the
  largest $\sigma_0^\ast$ such that $\Prob\b(\sigma_0)=0$ for all $\sigma_0\le \sigma_0^\ast$.
\end{enumerate}

We note that, in contrast to most of the other fiber-bundle models,
our stochastic models neglect any spatial correlations in the load
transfer after a failure. The only other model, as far as we are aware
of, that also uses a stochastic (rather than a spatially correlated)
stress redistribution is that of Dalton et al.~\cite{Dalton2009}, where it is
assumed that the load of a failing fiber is transferred to a fixed
number ($n = 1, 2, \dots$) of randomly chosen surviving fibers.

In addition, we use strength-threshold distributions that are truncated below
the initially applied stress~$\sigma_0$---in contrast to most models
studied in the fiber-bundle literature. There exist, however, a
number of investigations that also consider truncated threshold
distributions~\cite{Bhattacharyya2003,Raischel2006,Pradhan2008}, so
that a direct comparison with our results can be made.

In Sect.~\ref{sec:model}, we introduce our stochastic
load-redistribution model and describe its application to fracture
processes. Subsequently, in Sect.~\ref{sec:branching}, we describe a Markov
approximation of the model which leads to a description in terms of
generalized branching processes.  In Sects.~\ref{sec:rdld} and \ref{sec:pld}, we
analyze the two different model variants mentioned above, and final
conclusions are given in Sect.~\ref{sec:conclusions}.

\section{Stochastic load-redistribution model}
\label{sec:model}

\subsection{Load-redistribution rule and cascade model}

We shall consider a bundle consisting of $N$ fibers subjected to an
external force~$F$. We assume that the initial stress $\sigma_0 =
F/N$ of all fibers is equal and that the strength
thresholds~$\sigma\c$ of the individual fibers are randomly
distributed according to a probability density $\Probdc(\sigma\c)$.
For our setup of fracture-propagation experiments, we assume that we
start from a finite stress $\sigma_0>0$ and that all fibers with
thresholds smaller than~$\sigma_0$ have been
removed~\cite{Pradhan2008}. Thus, the threshold distribution has to
fulfill~$\Probdc(\sigma\c)=0$ for all $\sigma\c<\sigma_0$. On the
other hand, we are interested in a situation where already an
infinitesimal increase of the external force leads---with probability
one in the limit $N\to\infty$---to the breaking of exactly one fiber,
and we thus require $\Probdc(\sigma\c{=}\sigma_0)>0$.

When a fiber breaks, its stress has to be taken over by the remaining
intact fibers of the bundle. In our stochastic load-redistribution
model~\cite{Lehmann2009}, we assume that this process can be described by a rule of the form
\begin{equation}
  \sigma \to \sigma' = \sigma + \sigma\f \,\Delta\,.
  \label{eq:ssr}
\end{equation}
Here, $\sigma$ ($\sigma'$) is the stress of an intact fiber before
(after) the failure of a fiber with stress $\sigma\f$, and the
load-redistribution factor~$\Delta$ is a random number drawn
independently from the same distribution~$\Probd_\Delta(\Delta)$ for
each of the intact fibers. Note that for the initial failure,
both $\sigma$ and $\sigma\f$ are given by the initial
stress~$\sigma_0$.  In the special case~$\Delta=1/(N-1)$ of a uniform
load-redistribution, the form~\eqref{eq:ssr} reduces exactly to a GLS
rule. For a general non-uniform load redistribution, we require that
the failed stress will be shared \textit{on average} by the remaining
intact elements, i.e., the mean of $\Delta$ has to fulfill the condition
\begin{equation}
  \label{eq:redist_req}
  \langle \Delta \rangle = \frac{1}{N-1}\,.
\end{equation}

Due to the stress increment a fiber has obtained after a failure
event, its stress itself might be above its critical threshold. In
general, the initially failing fiber might thus induce the failure of
$N\f^{(1)}\ge1$ other fibers, thereby starting a failure cascade. We then
assume that all overloaded fibers fail simultaneously and that their
stress is again redistributed according to the rule~\eqref{eq:ssr},
where now both the pre-failure stresses~$\sigma$ of each of the
$N-N\f^{(1)}$ intact fibers and the stresses~$\sigma\f$ of each of the
$N\f^{(1)}$ failing fibers will, in general, be different random
variables. If this process leads to the further overloading of
$N\f^{(2)}\ge1$ fibers, it continues to a next cascade stage, and so
on. Eventually, either the bundle stabilizes again, i.e., all fibers
are stressed below their respective strength thresholds, or it breaks down
completely, i.e., all $N$ fibers fail. 

Note that, in general, during the failure cascade, the load
redistribution and hence the $\Delta$-distribution will be
modified. Whereas the details of such a modification, which can depend
on topological changes, are very difficult to model, one at least has
to take into account one dominant effect: As the number of
fibers~$N^{(s)}_\mathrm{in}$ that are still intact at cascade
stage~$s$ decreases, the mean~$\langle \Delta \rangle$ has to increase
in accordance with Eq.~\eqref{eq:redist_req} with $N$ replaced
by~$N^{(s)}_\mathrm{in}$. Below, we will discuss how to fulfill this
requirement for the chosen forms of load redistribution.

\subsection{Truncated strength-threshold distributions}

For our setup of fracture-propagation experiments, we have to truncate
the distribution of strength thresholds below $\sigma\c=\sigma_0$.
In the literature on fiber-bundle models, the strength
thresholds~$\sigma\c$ of the individual fibers are usually assumed to
be distributed according to a Weibull distribution with density $
\weibullexp\, \sigma\c^{\weibullexp-1}\exp(-\sigma\c^\weibullexp)$,
where mostly the Weibull index~$\weibullexp=2$ is used. Truncation then leads to a
distribution of the form
\begin{equation}
  \label{eq:weibull}
  \Probdc(\sigma\c) =
  \begin{cases}
    \weibullexp\, \sigma\c^{\weibullexp-1}\,\exp(\sigma_0^\weibullexp-\sigma\c^\weibullexp)& \text{if $\sigma_0\le\sigma\c$}\\
    0 & \text{else.}
  \end{cases}
\end{equation}
In addition, we shall also consider uniform distributions that are
truncated below the initial stress~$\sigma_0$:
\begin{equation}
  \label{eq:uniform}
  \Probdc(\sigma\c) =
  \begin{cases}
    \displaystyle\frac{1}{1-\sigma_0} & \text{if $\sigma_0\le\sigma\c\le1$}\\
    0 & \text{else.}
  \end{cases}
\end{equation}

In Sects.~\ref{sec:rdld} and \ref{sec:pld}, we summarize and discuss
the corresponding results for $\Prob\nc(\sigma_0)$,
$\Prob\b(\sigma_0)$ and $\sigma_0\crit$ for two different
load-redistribution models and for the two threshold
distributions~\eqref{eq:weibull} and \eqref{eq:uniform}.

\section{Generalized-branching-process approximation}
\label{sec:branching}

The dynamics of the stochastic cascade model described in the previous
section and the quantities $\Prob\nc(\sigma_0)$, $\Prob\b(\sigma_0)$
and $\sigma_0\crit$ can only be obtained exactly by means of
Monte-Carlo simulations. In the limit of large system sizes
$N\to\infty$, however, we can achieve an approximate description of
the cascade dynamics by noting the following points:
\begin{enumerate}[(i)]
\item The failure of a fiber leaves the stress in the majority of the
  intact fibers nearly unaffected, i.e., maximally leads to changes of the order
  of $1/N$.  Thus, along the failure cascade, the stress of the intact
  fibers is approximately given by the initial stress $\sigma_0$.

\item The remaining number of fibers always stays infinitely large and
  thus the number of further failures induced by a failing fiber is
  distributed according to a Poisson distribution.

\item The interaction between different failures can be neglected,
  i.e., in the case of several induced failures, the failure cascades
  resulting from each of these failures can be treated as being independent.
\end{enumerate}
Under these assumptions, the cascade dynamics becomes
Markovian~\cite{Hnggi1982} if we choose the point process of the
stresses of the failed fibers as underlying state space. This point
process on the semi-infinite interval $[\sigma_0,\infty)$ is
independent~\cite{van_Kampen2007}, and the failure dynamics can thus
be described by a generalized branching
process~\cite{Harris1963} with characteristic functional
\begin{equation}
  \begin{split}
  G[u;\sigma\f] = \exp\bigg\{\mu(\sigma\f)\bigg[ \int\! & \d\sigma\f' \, \, \Probd(\sigma\f'|\sigma\f'\,{>}\, \sigma\c;\sigma\f) 
  \,\e^{-u(\sigma\f')} - 1\bigg]\bigg\}
  \end{split}
  \label{eq:chfunc}
\end{equation}
for the point process induced by a single failure with
stress~$\sigma\f$.  Here, $\Probd(\sigma\f'|\sigma\f'>
\sigma\c;\sigma\f)$ denotes the conditional probability density that
the induced failure resulting from the breaking of a fiber with
stress~$\sigma\f$ occurs with a stress~$\sigma\f'$. For given
distributions of the load-redistribution factors~$\Delta$ and the
critical thresholds $\sigma\c$, this quantity can be readily
calculated from Eq.~\eqref{eq:ssr}. This also holds true for the mean
number of failures,
\begin{equation}
  \label{eq:muf}
 \muf(\sigma\f) = (N-1)\, \Prob(\sigma\f'>\sigma\c|\sigma\f) \,,
\end{equation}
induced by the breaking of a fiber with stress~$\sigma\f$. We remark
that in order for this quantity to be finite in the
limit~$N\to\infty$, the
probability~$\Prob(\sigma\f'>\sigma\c|\sigma\f)$ for the induced
failure of a given intact fiber has to vanish as $1/N$. This is in
accordance with the requirement~\eqref{eq:redist_req} for the mean of
the load-redistribution factors~$\Delta$. Finally, in
Eq.~\eqref{eq:chfunc}, $u$ denotes an arbitrary non-negative test function on the
interval~$[\sigma_0,\infty)$.

For later use, we note that from the mean number of induced failures,
one directly obtains the no-cascade probability, which is given by
\begin{equation}
  \Prob\nc(\sigma_0) = [1-P(\sigma\f'>\sigma\c|\sigma_0)]^{N-1}\,.
\end{equation}
In the limit~$N\to\infty$, this relation becomes
\begin{equation}
  \label{eq:pnc}
  \Prob\nc(\sigma_0) = \exp\big[-\muf(\sigma_0)\big]\,.
\end{equation}

In principle, the properties of the later cascade stages and thus the
full cascade dynamics can be obtained in a recursive way from the
functional~\eqref{eq:chfunc} (see Ref.~\cite{Harris1963}). Here, we
are only interested in the question whether an initial failure leads
to the breakdown of the entire fiber bundle. It can be
shown~\cite{Harris1963} that this question can be answered by
solving the integral equation
\begin{multline}
  \label{eq:pb}
  1-\Prob\b(\sigma\f) = \exp\bigg\{-\muf(\sigma\f)\int\!  \d\sigma\f' \, \, \Probd(\sigma\f'|\sigma\f'> \sigma\c;\sigma\f)
  \\\times \Prob\b(\sigma\f')\bigg\}
\end{multline}
for the probability~$\Prob\b(\sigma\f)$ that an initial failure with
stress~$\sigma\f$ leads to the breakdown of the entire bundle.  Using
Eq.~\eqref{eq:muf} and the load-redistribution rule~\eqref{eq:ssr},
together with the fact that, within the approximation considered, the
pre-failure stress~$\sigma$ is just given by the initial
stress~$\sigma_0$, we can rewrite the integral equation~\eqref{eq:pb}
in the form
\begin{multline}
  \label{eq:pb2}
  1-\Prob\b(\sigma\f) = \exp\bigg\{-\frac{1}{\sigma\f}
  \int\limits_{\sigma_0}^{\infty}\!  \d\sigma\f' \, F\c(\sigma\f')  \,
  \tilde \Probd_\Delta\bigg(\frac{\sigma\f'-\sigma_0}{\sigma\f}\bigg)
  \\\times \Prob\b(\sigma\f')\bigg\}\,,
\end{multline}
where $\tilde \Probd_\Delta(\Delta) = \lim_{N\to\infty} N
\Probd_\Delta(\Delta)$.  Here, we have also used the relation
\begin{equation}
  \label{eq:condprobrel}
  \Probd(\sigma\f'|\sigma\f' \,{>}\, \sigma\c;\sigma\f)\,
  \Prob(\sigma\f' \,{>}\, \sigma\c|\sigma\f) = \Probd(\sigma\f';\sigma\f' \,{>}\,
  \sigma\c|\sigma\f)  
\end{equation}
and introduced the cumulative distribution
function $F\c(\sigma\c)=\int_{-\infty}^{\sigma\c}\mathrm{d}\sigma\c'
\Probdc(\sigma\c')$ corresponding to the threshold distribution
$\Probdc(\sigma\c)$. Under quite general
assumptions~\cite{Harris1963}, the unique solution of the integral
equations~\eqref{eq:pb} or \eqref{eq:pb2} can be found by means of an
iterative procedure starting from an arbitrary initial guess for
$\Prob\b(\sigma\f)$. The so obtained probability function can then be
evaluated at the initial stress~$\sigma_0$ to obtain the
probability~$\Prob\b(\sigma_0)$ for the breakdown of a fiber bundle in
the setup described in Sect.~\ref{sec:intro}.

We finally remark that the interpretation of Eq.~\eqref{eq:pb} becomes
clear if one writes the exponential on the right-hand side in the form
\begin{equation}
  \label{eq:pb3}
  1-\Prob\b(\sigma\f) = 
  \lim_{N\to\infty} \bigg[1-\int\!  \d\sigma\f' \,\Probd(\sigma\f';\sigma\f'\,{>}\, \sigma\c|\sigma\f)
\Prob\b(\sigma\f')\bigg]^{N}\,,
\end{equation}
where we have again used Eqs.~\eqref{eq:muf} and
\eqref{eq:condprobrel}. Hence, the probability that no breakdown occurs after the
failure of a fiber with stress~$\sigma\f$ is equal---in the limit
$N\to\infty$---to the probability that none of the induced failures with
stress~$\sigma\f'$ leads to a breakdown.

\section{Stochastic model for range-dependent load redistribution ($\gamma$-model)}
\label{sec:rdld}

Hidalgo et al.~\cite{Hidalgo2002} proposed a FBM where the stress
transfer after a failure decays with the distance~$r$ between the
failing fiber and the one affected by the failure as a power law
$Z/r^\gamma$. Here, $Z$ is a normalization factor which ensures that
the total load is conserved. They furthermore assumed that all fibers
are arranged on a two-dimensional square lattice. By varying the
exponent~$\gamma>0$, they were then able to study the transition
between a GLS rule (for $\gamma\to0$) and a LLS rule (for
$\gamma\to\infty$).  Note that due to the infinite range of the
power-law transfer function, the latter situation of a strictly local
load-transfer can only be achieved in an approximate sense. For the
case of Weibull-distributed strength thresholds, a Monte-Carlo
analysis of this range-dependent load-transfer model showed that for
an exponent $\gamma\lesssim 2$, the model behaves essentially as a FBM
with GLS rule and, in particular, a finite critical stress value was
observed. For larger~$\gamma$, there is a transition to the LLS case
with a critical stress that vanishes in the large system-size limit.
In Ref.~\cite{Raischel2006}, the same model has been analyzed for
uniform threshold distributions with a lower cutoff
$\sigma_\mathrm{L}$, and in this case, the critical stress remains
finite for all values of $\gamma$ (if $\sigma_\mathrm{L} > 0$).

In the following, we study a stochastic version of this model, which
we shall call ``$\gamma$-model''. It is based on the assumption that the
position of the fibers is uniformly distributed within the
two-dimensional cross-section of the bundle. Upon failure of a fiber
we then randomly pick the affected fibers from this uniform
distribution and calculate the load-transfer factor~$\Delta$ according
to the (random) distance. We will now first derive the corresponding
$\Delta$-distribution, then analyze the properties of the resulting
model, and finally compare its results to the ones obtained in
Ref.~\cite{Raischel2006}.

\subsection{Distribution of load-redistribution factors}

For reasons of simplicity, we assume that the broken fiber is in the
center of a hollow cylinder with inner (outer) radius~$\rmin$
($\rmax$) containing $N-1$~intact fibers uniformly distributed with
area density~$\varrho$ in the cross-sectional area of size $A = \pi
(\rmax^2-\rmin^2)$. Note that in contrast to
Refs.~\cite{Raischel2006,Hidalgo2002}, we consider a uniform
distribution of the fiber positions and thus have to introduce a lower
cut-off for the distance~$r$ to prevent a divergence at small
distances.

For this uniform spatial distribution and the given distance
dependence of the load transfer, we can then readily derive the
probability distribution for the load-redistribution factors~$\Delta$ appearing
in Eq.~\eqref{eq:ssr}:
\begin{align}
  \label{eq:Delta_r_gamma}
  \Probd_\Delta(\Delta) & {}:= \frac{2\pi}{A} \int\limits_\rmin^\rmax \!\d r \,r\, \delta(\Delta - Z/r^\gamma)\\
  & {} = 
  \label{eq:Delta_r_gamma2}
  \begin{cases}
    \displaystyle \frac{2\pi\,Z^{\frac{2}{\gamma}} \, \Delta^{-\frac{\gamma+2}{\gamma}}}{A\gamma}  & \text{for $\,\Delta_\mathrm{min}\le\Delta\le\Delta_\mathrm{max}$}\\
    0 & \text{otherwise.}
  \end{cases}
\end{align}
Here, the lower and upper cutoffs for $\Delta$ are given by
$\Delta_\mathrm{min} = Z / \rmax^\gamma$ and $\Delta_\mathrm{max} =
Z/\rmin^\gamma$, respectively.  In the stochastic model, we fix the
constant~$Z$ by imposing Eq.~\eqref{eq:redist_req}, which states that,
on average, the load is redistributed to the remaining
elements.  This yields the normalization constant
\begin{equation}
  \label{eq:Z}
  Z = \frac{2-\gamma}{2 (N-1)} \, \frac{\rmax^2-\rmin ^2}{\rmax^{2-\gamma}-\rmin^{2-\gamma}}
\end{equation}
for $\gamma\ne 2$; the special case~$\gamma=2$ can be readily treated by considering the limit~$\gamma\to2$.

We can write Eqs.~\eqref{eq:Delta_r_gamma2} and \eqref{eq:Z} in a more
convenient form by introducing a dimensionless length $L=\rmax/\rmin$. This first 
allows us to express the number of intact fibers as
\begin{equation}
\label{eq:N_vs_L}
  N-1 = s \left(L^2-1\right)\,,
\end{equation}
where $s = \pi \rmin^2 \, \varrho\approx1$ is the average number
of  fibers in the vicinity of the failing fiber. From
Eq.~\eqref{eq:N_vs_L}, we obtain, e.g., $s=\pi/4$ if we assume that
the model describes the continuous approximation of fibers located on
a quadratic lattice (with lattice constant $\rmin$) consisting of $N
\sim \pi (L/2)^2$ sites inside a circle of the given radius. Furthermore, we can write 
the $\Delta$-cutoffs in the form 
\begin{align}
  \label{eq:Delta_min}
  \Delta_\mathrm{min} & {} = \frac{1}{2s} \, \frac{1}{L^2}\,\frac{2-\gamma}{1-L^{\gamma-2}} \\
  \label{eq:Delta_max}
  \Delta_\mathrm{max} & {} = \frac{1}{2s} \, \frac{2-\gamma}{L^{\gamma-2}-1} = L^\gamma \, \Delta_\mathrm{min}\,,
\end{align}
where again the case $\gamma=2$ has to be treated as limit. The
probability distribution is then given by the power-law form
\begin{equation}
  \label{eq:Delta_r_gamma3}
  \Probd_\Delta(\Delta) =  \frac{2}{\gamma}\, \frac{1}{L^2-1}\,\frac{1}{\Delta_\mathrm{max}} \, 
  \left(\frac{\Delta}{\Delta_\mathrm{max}}\right)^{-\frac{\gamma+2}{\gamma}}
\end{equation}
for $\,\Delta_\mathrm{min}\le\Delta\le\Delta_\mathrm{max}$.

For later use, we note that in the limit of large system sizes, the
lower $\Delta$-cutoff always scales to zero: $\Delta_\mathrm{min}\to 0$ for
$L\to\infty$. The behavior of the upper cutoff, however, strongly depends on the
exponent~$\gamma$: For $\gamma\le2$, $\Delta_\mathrm{max}$ also vanishes in the limit $L\to\infty$, whereas
for $\gamma>2$, it converges to a finite value:
$\Delta_\mathrm{max}\to (\gamma-2)/2s$ for $L\to\infty$.

\subsection{Mean number of directly induced fiber failures}
\label{sec:power_law_mean}

We now calculate the mean number~$\muf(\sigma\f)$ of fiber failures resulting
directly from the breaking of a fiber with
stress~$\sigma\f$. From rule~\eqref{eq:ssr} and Eq.~\eqref{eq:muf}, we obtain
\begin{equation}
  \label{eq:mu_f_gamma}
  \begin{split}
  \muf(\sigma\f) & {}= (N-1) \Prob(\sigma_0 + \sigma\f\, \Delta\,{>}\,\sigma\c)    \\
  & {} = (N-1) \int \d\Delta \, \Probd_\Delta(\Delta) \, F\c(\sigma_0 + \sigma\f \, \Delta)\,.
  \end{split}
\end{equation}
With Eqs.~\eqref{eq:N_vs_L}--\eqref{eq:Delta_r_gamma3}, we
can write this expression in the form
\begin{equation}
  \label{eq:mu_f_gamma2}
  \muf(\sigma\f) = \frac{2s}{\gamma} \int\limits_{L^{-\gamma}}^1 \d x\, x^{-\frac{\gamma+2}{\gamma}} \, 
  F\c(\sigma_0 + \sigma\f \, \Delta_\mathrm{max} \, x)\,.
\end{equation}
For the evaluation of this integral in the limit $L\to\infty$, where
the integrand becomes singular, it is useful to consider the cases
$\gamma\le2$ and $\gamma>2$ separately.

(i) $\gamma\le 2$: As mentioned above, we then have $\Delta_\mathrm{max}\to0$ for $L\to\infty$ and thus
consider the Taylor expansion 
\begin{equation}
  \label{eq:Fc_taylor}
 F\c(\sigma_0 + \sigma\f \, \Delta_\mathrm{max} \, x) = 
 \sum_{l=1}^\infty \frac{(\sigma\f \,\Delta_\mathrm{max} \, x)^l}{l!} \, F\c^{(l)}(\sigma_0)\,,
\end{equation}
where we have used that $F\c(\sigma_0)=0$.
Inserting this expansion into Eq.~\eqref{eq:mu_f_gamma2}, we find with
Eq.~\eqref{eq:Delta_max} that only the first order term~$l{=}1$ leads
to a non-vanishing contribution in the limit $L\to\infty$. This yields
\begin{equation}
  \label{eq:mu_f_gamma3a}
     \muf(\sigma\f) \to  \sigma\f \, \Probdc(\sigma_0)\,.
\end{equation}
Note that this result is independent of the exponent~$\gamma$.

(ii) $\gamma>2$: In this case, we can insert the finite asymptotic
value for $\Delta_\mathrm{max}$, $\Delta_\mathrm{max} = (\gamma - 2) /
2s$ for $L \to\infty$, into Eq.~\eqref{eq:mu_f_gamma2} and write the
mean number of failures as the improper integral
\begin{equation}
  \label{eq:mu_f_gamma3b}
  \muf(\sigma\f) = \frac{2s}{\gamma} \int\limits_{0}^1 \d x\, x^{-\frac{\gamma+2}{\gamma}} \, 
  F\c(\sigma_0 + \sigma\f \, (\gamma-2)/(2s)\, x)\,.
\end{equation}
Here, the convergence of the integral at the lower boundary is
guaranteed since, because the strength-threshold distribution is
truncated below~$\sigma_0$, we have $F\c(\sigma_0 + \sigma\f \,
(\gamma-2)/(2s)\, x)=\mathcal{O}(x)$.

\subsection{Breakdown probability}
\label{sec:power_law_pb}

For the evaluation of the breakdown probability, we have to solve the integral equation~\eqref{eq:pb2}, which in
the present case assumes the form
\begin{multline}
  1-\Prob\b(\sigma\f) =
  \exp\Bigg\{
    -\frac{2s}{\gamma\sigma\f\,\Delta_\mathrm{max}}\!\!
    \int\limits_{\sigma_0 + \sigma\f \Delta_\mathrm{min}}^{\sigma_0 + \sigma\f \Delta_\mathrm{max}}
    \!\!\!\!
    \d \sigma\f' \,
    \left(
      \frac{\sigma\f'-\sigma_0}{\sigma\f\,\Delta_\mathrm{max}}
    \right)^{-\frac{\gamma+2}{\gamma}}\\
     \times F\c(\sigma\f')\, \Prob\b(\sigma\f')
  \Bigg\}\,.
  \label{eq:pb_iter_gamma}
\end{multline}
Alternatively, we can write this equation as
\begin{multline}
  1-\Prob\b(\sigma\f) =
  \exp\Bigg\{
    -\frac{2s}{\gamma}
    \int\limits_{L^{-\gamma}}^{1}
    \d x\,
    x^{-\frac{\gamma+2}{\gamma}}
     F\c(\sigma_0 + \sigma\f \Delta_\mathrm{max} x)\, \\
     \times \Prob\b(\sigma_0 + \sigma\f \Delta_\mathrm{max} x)
  \Bigg\}\,.
  \label{eq:pb_iter_gamma2}
\end{multline}
The integral on the right-hand side of this equation has to be
evaluated in the limit of infinitely large systems ($L\to\infty$).
Again, we treat the cases $\gamma\le2$ and $\gamma>2$ separately.

(i) $\gamma\le2$: As above, in Eq.~\eqref{eq:Fc_taylor}, we expand the
distribution function $F\c(\sigma\c)$ around $\sigma\c=\sigma_0$ and now furthermore
assume that such an expansion is also valid for the breakdown
probability,
\begin{equation}
 \Prob\b(\sigma_0 + \sigma\f \Delta_\mathrm{max} x) = 
 \sum_{m=0}^\infty \frac{(\sigma\f \,\Delta_\mathrm{max} \, x)^m}{m!} \, \Prob\b^{(m)}(\sigma_0)\,.
\end{equation}
Inserting these expansions into the integral in
Eq.~\eqref{eq:pb_iter_gamma2}, we find with Eq.~\eqref{eq:Delta_max}
that the various terms behave as $L^{2(1-l-m)}$ for large $L$. Thus,
only the lowest order terms $l=1$ and $m=0$ survive in this limit. The
integral equation~\eqref{eq:pb_iter_gamma2} hence simplifies to 
the transcendental equation
\begin{equation}
  1-\Prob\b(\sigma\f) =
  \exp\big[
    -\sigma\f\,
    \Probdc(\sigma_0)
    \, \Prob\b(\sigma_0)
    \big]
    \,,
  \label{eq:pb_iter_gamma3a}
\end{equation}
which, again, is $\gamma$-independent. In this regime, the critical
stress~$\sigma_0\crit$ is given by the condition that the mean number
of directly induced failures equals unity:
\begin{equation}
  \label{eq:sigma0crit_power_law}
  \mu(\sigma_0\crit)=\sigma_0\crit \, \Probd\c(\sigma_0\crit) = 1
\end{equation}

(ii) $\gamma>2$: As for the evaluation of the mean number of induced
failures, cf.\ Eq.~\eqref{eq:mu_f_gamma3b}, we use the asymptotic
value of $\Delta_\mathrm{max}$ and replace the integral in the limit
$L\to\infty$ by an improper one [cf.\ also remark after
Eq.~\eqref{eq:mu_f_gamma3b}].  This leads to the integral equation
\begin{multline}
  1-\Prob\b(\sigma\f) =
  \exp\Bigg\{
    -\frac{2s}{\gamma}
    \int\limits_{0}^{1}
    \d x\,
    x^{-\frac{\gamma+2}{\gamma}}
     F\c(\sigma_0 + \sigma\f (\gamma-2)/(2s) x)\, \\
     \times \Prob\b(\sigma_0 + \sigma\f (\gamma-2)/(2s)x)
  \Bigg\}\,,
  \label{eq:pb_iter_gamma3b}
\end{multline}
which, in general, can only be solved numerically, e.g., by means of an iterative procedure. In particular, 
the critical stress~$\sigma_0\crit$ is not determined by a simple relation like Eq.~\eqref{eq:sigma0crit_power_law} but
has to be determined from the full solution of Eq.~\eqref{eq:pb_iter_gamma3b}.

\subsection{Results}
\label{sec:results_power_law}

\subsubsection{Uniform distribution of strength thresholds}
\label{sec:results_power_law_uniform}

In the case of a uniform distribution~\eqref{eq:uniform} of the
strength thresholds, the mean number of induced
failures~\eqref{eq:mu_f_gamma2} can be evaluated explicitly. With
$\sigma\f=\sigma_0$ we then obtain from Eq.~\eqref{eq:pnc} the
no-cascade probability
\begin{subequations}
  \label{eq:mu_power_law_uniform}
  \begin{equation}
    \Prob\nc(\sigma_0) = 
    \exp\bigg(-\frac{\sigma_0}{1-\sigma_0}\bigg) 
   \end{equation}
if $\gamma<2$ or $\sigma_0\le [1+(\gamma-2)/(2s)]^{-1}$, and
 \begin{equation}
   \Prob\nc(\sigma_0) = 
   \exp\bigg[s-\frac{\gamma}{2} \left(\frac{\gamma-2}{2s}\right)^{2/\gamma-1}\left(\frac{\sigma_0}{1-\sigma_0}\right)^{2/\gamma} \bigg]
 \end{equation}
otherwise.
\end{subequations}
For the calculation of the breakdown probability~$\Prob\b(\sigma_0)$,
the transcendental equation~\eqref{eq:pb_iter_gamma3a} (for $\gamma\le
2$) or the integral equation~\eqref{eq:pb_iter_gamma3b} (for
$\gamma>2$) have to be solved numerically.

\begin{figure}[t]
  \centering
  \includegraphics[width=0.9\columnwidth]{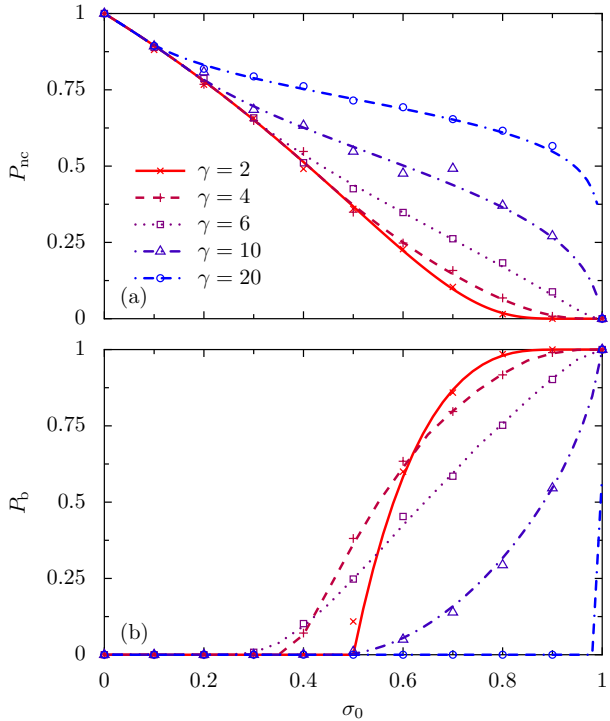}
  \caption{$\gamma$-model with $s=\pi/4$ and uniform distribution
    of strength thresholds. (a) No-cascade probability $\Prob\nc$ and (b) breakdown
    probability $\Prob\b$ as a function of the initial
    stress~$\sigma_0$ for different values of the exponents~$\gamma$.
    Lines: Results from Eq.~\eqref{eq:mu_power_law_uniform} [panel (a)] and
    Eqs.~\eqref{eq:pb_iter_gamma3a} and \eqref{eq:pb_iter_gamma3b}
    [panel (b)], respectively. Symbols: Results from Monte-Carlo
    simulations of the failure process for $L=64$ averaged over $1000$
    realizations. The statistical error is of the order of the size of
    the symbols.}
  \label{fig:pnc_pb_power_law_uniform_sigma0}
\end{figure}
In Fig.~\ref{fig:pnc_pb_power_law_uniform_sigma0} we show the
no-cascade probability~$\Prob\nc(\sigma_0)$ and the breakdown
probability~$\Prob\b(\sigma_0)$ as a function of the initial
stress~$\sigma_0$ for different values of the exponent~$\gamma$ in
Eq.~\eqref{eq:Delta_r_gamma3}. The approximate results within the
Markov approximation (lines) are compared with Monte-Carlo simulations
(symbols) of the failure dynamics generated by the load
redistribution~\eqref{eq:ssr}. We note that in the Monte-Carlo
simulations, we have to ensure that condition~\eqref{eq:redist_req} stays
fulfilled during the entire cascade process. This is done by replacing
$L$ in Eqs.~\eqref{eq:Delta_min}--\eqref{eq:Delta_r_gamma3} by
$L_\mathrm{eff} = \sqrt{N_\mathrm{in}/s +1}$,
where $N_\mathrm{in}$ denotes the number of remaining intact
fibers.

\begin{figure}[t]
  \centering
  \includegraphics[width=0.9\columnwidth]{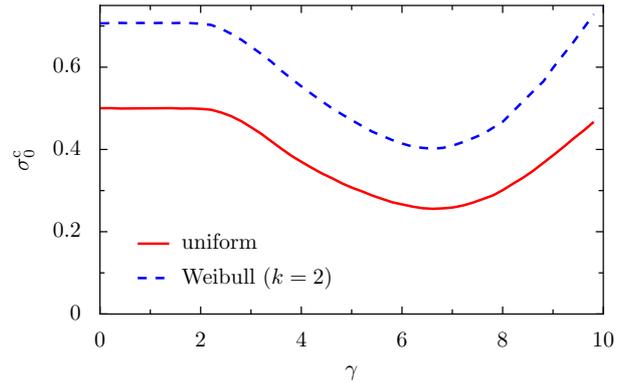}
  \caption{$\gamma$-model with $s=\pi/4$. Critical stress
    $\sigma_0\crit$ as a function of the exponent~$\gamma$. Solid
    line: uniformly distributed strength thresholds. Dashed line:
    Weibull-distributed ($\weibullexp=2$) strength thresholds.}
  \label{fig:sigma0crit_power_law}
\end{figure}
With increasing initial stress~$\sigma_0$, we observe a gradual
decrease of the no-cascade probability from one to zero [cf.\
Fig.~\ref{fig:pnc_pb_power_law_uniform_sigma0}(a)]. In contrast, the
breakdown probability
[Fig.~\ref{fig:pnc_pb_power_law_uniform_sigma0}(b)] exhibits a
critical behavior: There is a $\gamma$-dependent critical
stress~$\sigma_0\crit$ such that for $\sigma_0 \le
\sigma_0\crit$, the probability of a breakdown of the fiber
bundle vanishes exactly. The dependence of the critical stress on the
exponent~$\gamma$ is illustrated in Fig.~\ref{fig:sigma0crit_power_law}. The
value~$\sigma_{0}\crit=1/2$ for $\gamma\le2$ can be readily determined
from Eqs.~\eqref{eq:uniform} and \eqref{eq:sigma0crit_power_law} and exactly reproduces the value of the FBM with GLS
rule~\cite{Kun2000}.  For $\gamma>2$, and hence smaller
effective ``range'' of the stress redistribution, we first observe a
transition to a regime, where the critical stress decreases with
increasing~$\gamma$ down to a minimal value
$\sigma_0\crit\approx0.26$.  For even larger~$\gamma$, $\gamma\gtrsim7$,
the critical stress increases again.  We remark, however, that for
large~$\gamma$, it becomes numerically rather difficult to find the
precise location of the critical transition because the onset of the
regime with a finite breakdown probability becomes more and more flat
[cf.\ Fig.~\ref{fig:pnc_pb_power_law_uniform_sigma0}(b)]. 

It is interesting to compare our results with the ones obtained for
the variable-range load-redistribution model of
Ref.~\cite{Raischel2006}, in particular, the behavior for the case of
the failure stress being equal to the cutoff strength [cf.\ Fig.~1(a)
of Ref.~\cite{Raischel2006}]. In both models, we observe a critical
value of $\gamma\approx 2$ above which a transition from a GLS regime
to one with short-ranged stress transfer and smaller
$\sigma_0\crit$-value takes place. Within our model, we can trace back
this transition to a change in the load-redistribution distribution,
in particular, the asymptotic value of
$\Delta_\mathrm{max}$. Furthermore, the critical stress values of both
models agree rather well up to $\gamma\lesssim7$.  For even larger
exponents~$\gamma$, we find an increase of the critical stress, which
cannot be observed in the more microscopic model of
Ref.~\cite{Raischel2006}. This discrepancy probably results from a
breakdown of the continuum approximation upon which the
distribution~\eqref{eq:Delta_r_gamma} is based. The deficiency of this
approximation for large~$\gamma$ is also reflected in the fact that
for $\gamma>2+2s$, the asymptotic value for $\Delta_\mathrm{max}$
becomes larger than unity, which means that a single fiber may receive
a stress increment that is higher than the stress of the failing
fiber. In order to prevent such a pathological behavior, a more
sophisticated load-redistribution model has to be used.

We finally note that the results from the Markov approximation, i.e.,
the generalized branching process description, agree very well with
the ones obtained by Monte-Carlo simulations of the failure process.
Around the critical transition, some deviations for the breakdown
probability can be observed, which, however, decrease with increasing
system size~$N$ and, thus, represent finite-size effects~\cite{Lehmann2009}.

\subsubsection{Weibull distribution of strength  thresholds}

For the case of the truncated Weibull distribution~\eqref{eq:weibull}
of strength thresholds, we have to evaluate for $\gamma>2$ both the
no-cascade probability and the breakdown probability numerically from
Eqs.\eqref{eq:mu_f_gamma3b} and \eqref{eq:pb_iter_gamma3b},
respectively.

\begin{figure}[t]
  \centering
  \includegraphics[width=0.9\columnwidth]{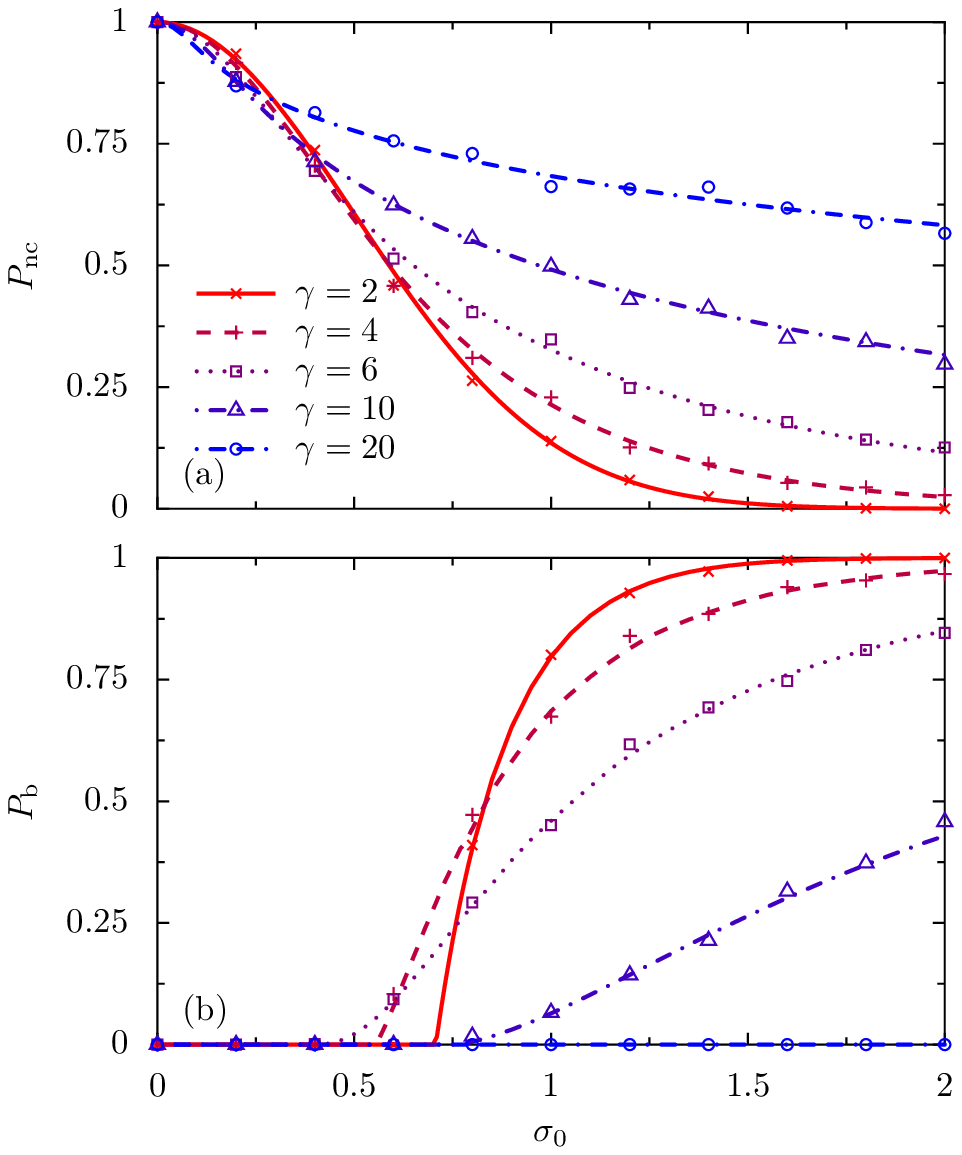}
  \caption{$\gamma$-model with $s=\pi/4$ and Weibull-distributed
    strength thresholds ($\weibullexp=2$). (a) No-cascade probability $\Prob\nc$ and (b) breakdown
    probability $\Prob\b$ as a function of the initial
    stress~$\sigma_0$ for different values of the exponents~$\gamma$.  Lines: Results from
    Eqs.~\eqref{eq:mu_f_gamma3a} and \eqref{eq:mu_f_gamma3b} [panel (a)] and
    Eqs.~\eqref{eq:pb_iter_gamma3a} and \eqref{eq:pb_iter_gamma3b}
    [panel (b)], respectively. Symbols: Results from Monte-Carlo
    simulations of the failure process for $L=64$ averaged over $1000$
    realizations. The statistical error is of the order of the size of
    the symbols.}
  \label{fig:pnc_pb_power_law_weibull_sigma0}
\end{figure}

The results as a function of the initial stress~$\sigma_0$ are
depicted in Fig.~\ref{fig:pnc_pb_power_law_weibull_sigma0}, where we
have chosen here and in the following a Weibull
index of~$\weibullexp=2$. Comparing with the case of a uniform
distribution of the strength thresholds, we find qualitatively the
same behavior. In particular, we identify a critical transition at a
$\gamma$-dependent stress~$\sigma_0\crit$ (cf.\
Fig.~\ref{fig:sigma0crit_power_law}), and again, for~$\gamma\le 2$,
the result $\sigma_{0}\crit=(1/k)^{1/k}$ for the GLS
case~\cite{Daniels1945} is recovered exactly from
Eq.~\eqref{eq:sigma0crit_power_law} together with the
distribution~\eqref{eq:weibull}.

\begin{figure}[t]
  \centering
  \includegraphics[width=0.9\columnwidth]{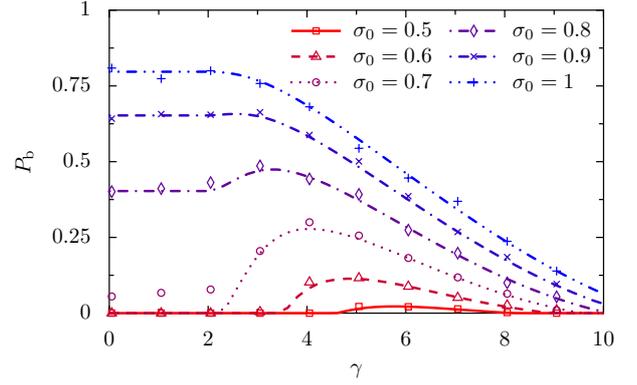}
  \caption{$\gamma$-model with $s=\pi/4$ and Weibull-distributed
    strength thresholds ($\weibullexp=2$). Breakdown probability
    $\Prob\b$ as a function of the exponent~$\gamma$ for different
    values of the initial stress~$\sigma_0$.  Lines: Results from
    Eqs.~\eqref{eq:pb_iter_gamma3a} and
    \eqref{eq:pb_iter_gamma3b}. Symbols: Results from Monte-Carlo
    simulations of the failure process for $L=64$ averaged over $1000$
    realizations. The statistical error is of the order of the size of
    the symbols.}
  \label{fig:pb_power_law_weibull_gamma}
\end{figure}
Figure~\ref{fig:pb_power_law_weibull_gamma} shows the dependence of
the breakdown probability on the exponent~$\gamma$ for fixed initial
stress~$\sigma_0$.  In accordance with Eq.~\eqref{eq:pb_iter_gamma3a},
the breakdown probability is $\gamma$-independent for $\gamma\le2$ and
assumes the GLS value.  In the case~$\gamma>2$, we find for
$\sigma_0\gtrsim1$ a regime with a monotonic decrease of $\Prob\b$
towards zero as a function of~$\gamma$. For smaller $\sigma_0$, the
breakdown probability assumes a maximum at a certain $\gamma$-value
and then decreases again towards zero. Finally, for $\sigma_0$ smaller
than the critical stress of the GLS model but larger than the minimal
stress observed in Fig.~\ref{fig:sigma0crit_power_law}, an increase of
$\gamma$ eventually leads to a destabilization of the system, i.e., a
non-vanishing breakdown probability, above some critical $\gamma$-value.

\section{Simple bimodal load-redistribution model ($\Delta_0$-model)}
\label{sec:pld}

In Ref.~\cite{Lehmann2009}, we have introduced a simple prototype
model that interpolates between the limiting cases of global load
redistribution and the transfer of the failing load to a single other
element.

The model, which we shall call ``$\Delta_0$-model'', is characterized
by a bimodal distribution of the load-redistribution factors~$\Delta$,
\begin{equation}
  \Delta = 
  \begin{cases}
    \Delta_0 & \text{with probability $p_0$} \\
    0 & \text{with probability $1-p_0$}\,,
  \end{cases}
  \label{eq:bimodal}
\end{equation}
i.e., after the failure of an element with stress~$\sigma\f$, the
stress~$\sigma$ of a still intact element is increased to $\sigma' =
\sigma +\Delta_0 \,\sigma\f$ with probability~$p_0$ and remains
unchanged with probability~$1-p_0$. 

We further require that the sum of the induced stress increments is, on average, equal to
the stress of the failing element and that~$\Delta_0\le 1$. It follows that 
\begin{equation}
  p_0 \, \Delta_0 = \frac{1}{N-1}\,,\quad \frac{1}{N-1}\le\Delta_0\le 1\,.
\end{equation}
$\Delta_0 = 1/(N-1)$ then corresponds to the limiting case of global
stress redistribution and $\Delta_0=1$ to the case where the failing
load is transferred, on average, to a single other element.

The probability that after the failure of a fiber with
stress~$\sigma\f$, a still intact fiber also fails can be written as
\begin{equation}
  p_0 \,\Prob(\sigma\c\,{<}\,\sigma_0 + \Delta_0 \sigma\f) 
  = 
  \frac{1}{(N-1) \Delta_0} \, \Prob(\sigma\c\,{<}\,\sigma_0 + \Delta_0\sigma\f)    
\end{equation}
and the mean number of induced failures becomes
\begin{equation}
  \label{eq:mu_bimodal}
  \mu(\sigma\f) = \frac{1}{\Delta_0}\, \Prob(\sigma\c \,{<}\, \sigma_0 +\Delta_0 \sigma\f)\,.
\end{equation}
In these expressions, we have neglected that a fraction of the still
intact fibers at later cascade stages may carry a stress larger
than~$\sigma_0$. It can be shown, however, that this is a finite-size
effect, i.e., this fraction vanishes as $N\to\infty$~\cite{Lehmann2009}.

The no-cascade probability then follows directly from
Eq.~\eqref{eq:pnc} by using Eq.~\eqref{eq:mu_bimodal} with
$\sigma\f=\sigma_0$:
\begin{equation}
\Prob\nc(\sigma_0) = \exp\bigg[-\frac{1}{\Delta_0} \, \Prob (\sigma\c \,{<}\, \sigma_0(1+\Delta_0))\bigg]\,.
\end{equation}
To calculate the breakdown probability~$\Prob\b(\sigma_0)$, we use
Eq.~\eqref{eq:pb2}, which reduces in the present case to the recursion
relation
\begin{equation}
  \label{eq:recursion_pb_bimodal}
  \Prob\b(\sigma_n) = 1 - \exp[-\mu(\sigma_n) \, \Prob\b(\sigma_{n+1})]\,,
\end{equation}
where
\begin{equation}
  \label{eq:recursion_pb_bimodal2}
  \sigma_n = \sigma_0 \frac{1-\Delta_0^{n+1}}{1-\Delta_0}\,,\quad n=0,1,\dots
\end{equation}
This recursion can be solved numerically to an arbitrary degree of
accuracy by starting at a high enough value of~$n$,
say~$n_\mathrm{s}$, and setting~$\Prob\b(n_\mathrm{s}) = 1$.

Finally, it can be shown that the critical stress~$\sigma_0\crit$ is
determined by
\begin{equation}
  \mu(\sigma_{n\to\infty}) = 1\,,\quad \sigma_{n\to\infty} = \frac{\sigma_0}{1-\Delta_0}\,,
\end{equation}
i.e., by
\begin{equation}
  \frac{1}{\Delta_0}\, \Prob\bigg(\sigma_c-\sigma_0 < \frac{\Delta_0\,\sigma_0}{1-\Delta_0}\bigg) = 1\,.
\end{equation}

\subsection{Results}

\subsubsection{Uniform distribution of strength thresholds}
\label{sec:bimodal_uniform}

For a uniform threshold distribution~\eqref{eq:uniform} we obtain
\begin{equation}
  \Prob(\sigma_c\,{<}\,\sigma_0+\Delta_0\sigma\f) = \min\bigg(1, \frac{\Delta_0\,\sigma\f}{1-\sigma_0}\bigg)\,,
\end{equation}
and it follows that
\begin{equation}
  \Prob\nc(\sigma_0) = 
  \begin{cases}
    \displaystyle\exp\bigg(-\frac{\sigma_0}{1-\sigma_0}\bigg) &\text{if $\displaystyle\sigma_0\le\frac{1}{1+\Delta_0}$} \\
    \displaystyle\exp\bigg(-\frac{1}{\Delta_0}\bigg) & \text{otherwise}
  \end{cases}
\end{equation}
and
\begin{equation}
  \label{eq:sigma0crit_bimodal_uniform}
  \sigma_0\crit = \frac{1-\Delta_0}{2-\Delta_0}\,.
\end{equation}

The breakdown probability $\Prob\b(\sigma_0)$ is determined by
numerically solving the recursion defined in
Eqs.~\eqref{eq:recursion_pb_bimodal} and
\eqref{eq:recursion_pb_bimodal2} with
\begin{equation}
  \mu(\sigma_n) = \frac{1}{\Delta_0}\, \min\bigg(1, \frac{\Delta_0\,\sigma_n}{1-\sigma_0}\bigg)\,.
\end{equation}

\subsubsection{Weibull distribution of strength thresholds}

For the case of a truncated Weibull distribution of
strength thresholds, Eq.~\eqref{eq:weibull}, with Weibull
index~$\weibullexp=2$, we have
\begin{equation}
  \Prob(\sigma\c\,{<}\,\sigma_0+\Delta_0\,\sigma\f)
  = 1-\exp\big[-\Delta_0\,\sigma\f\,(2\sigma_0 + \Delta_0\sigma\f)\big]
\end{equation}
and obtain
\begin{equation}
  \label{eq:pnc_bimodal_weibull}
  \Prob\nc(\sigma_0) = 
  \exp\bigg\{-\frac{1}{\Delta_0}\, \big[1-\exp\big(-\Delta_0\,(2+\Delta_0)\,\sigma_0^2\big)\big]\bigg\}
\end{equation}
and
\begin{equation}
  \label{eq:sigma0crit_bimodal_weibull}
  \sigma_0\crit = (1-\Delta_0) \left[\frac{-\ln(1-\Delta_0)}{\Delta_0\,(2-\Delta_0)}\right]^{1/2}\,.
\end{equation}
The corresponding results for arbitrary Weibull indices can be readily obtained.
$\Prob\b(\sigma_0)$ is again determined numerically by solving the
recursion of Eqs.~\eqref{eq:recursion_pb_bimodal} and
\eqref{eq:recursion_pb_bimodal2}, with
\begin{equation}
  \label{eq:mu_bimodal_weibull}
  \mu(\sigma_n) = \frac{1}{\Delta_0} 
  \bigg\{
    1-\exp\big[ -\Delta_0\, \sigma_n \,(2\sigma_0+\Delta_0\, \sigma_n)\big]
  \bigg\}\,.
\end{equation}

\subsubsection{Discussion}

\begin{figure}[t]
  \centering
  \includegraphics[width=0.9\columnwidth]{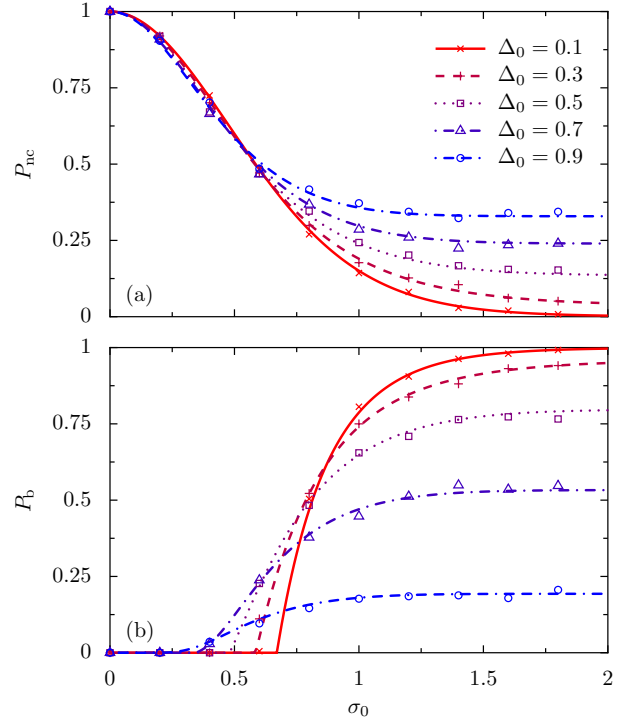}
  \caption{$\Delta_0$-model with Weibull-distributed strength
    thresholds ($\weibullexp=2$). (a) No-cascade probability
    $\Prob\nc$ and (b) breakdown probability $\Prob\b$ as a function
    of the initial stress~$\sigma_0$ for different values of the
    load-redistribution parameter~$\Delta_0$.  Lines: Results from
    Eqs.~\eqref{eq:pnc_bimodal_weibull} [panel (a)] and
    Eqs.~\eqref{eq:recursion_pb_bimodal},
    \eqref{eq:recursion_pb_bimodal2}, and
    \eqref{eq:mu_bimodal_weibull} [panel (b)], respectively. Symbols:
    Results from Monte-Carlo simulations of the failure process for
    $N=1000$ fibers averaged over $1000$ realizations. The statistical
    error is of the order of the size of the symbols.}
  \label{fig:pnc_pb_bimodal_weibull_sigma0}
\end{figure}

\begin{figure}[t]
  \centering
  \includegraphics[width=0.9\columnwidth]{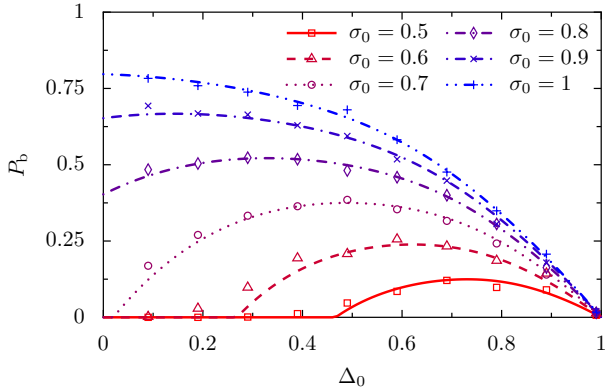}
  \caption{$\Delta_0$-model with Weibull-distributed strength
    thresholds ($\weibullexp=2$). Breakdown probability $\Prob\b$ as a function of the
    load-redistribution parameter~$\Delta_0$ for different values of
    the initial stress~$\sigma_0$. Lines: Results from
    Eqs.~\eqref{eq:recursion_pb_bimodal},
    \eqref{eq:recursion_pb_bimodal2}, and
    \eqref{eq:mu_bimodal_weibull}. Symbols: Results from Monte-Carlo
    simulations of the failure process for $N=1000$ fibers averaged over
    $1000$ realizations. The statistical error is of the order of the
    size of the symbols.}
  \label{fig:pb_bimodal_weibull_delta0}
\end{figure}

\begin{figure}[t]
  \centering
  \includegraphics[width=0.9\columnwidth]{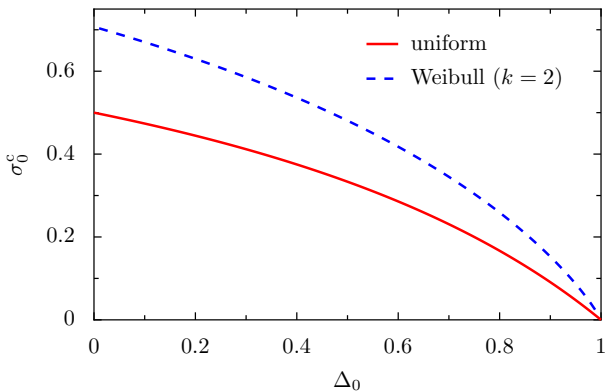}
  \caption{$\Delta_0$-model. Critical stress $\sigma_0\crit$ as a function of the
    load-redistribution parameter~$\Delta_0$. Solid
    line: uniformly distributed strength thresholds
    [Eq.~\eqref{eq:sigma0crit_bimodal_uniform}]. Dashed line:
    Weibull-distributed ($\weibullexp=2$) strength thresholds
    [Eq.~\eqref{eq:sigma0crit_bimodal_weibull}] .}
  \label{fig:sigma0crit_bimodal}
\end{figure}

The behavior of $\Prob\nc$ and $\Prob\b$ as a function of $\sigma_0$
and $\Delta_0$ is illustrated in
Figs.~\ref{fig:pnc_pb_bimodal_weibull_sigma0} and
\ref{fig:pb_bimodal_weibull_delta0} for a truncated Weibull
distribution of strength thresholds, and we note that the results for
uniformly distributed strength thresholds (see
Sect.~\ref{sec:bimodal_uniform}) show a qualitatively similar
behavior. Figure~\ref{fig:sigma0crit_bimodal} shows the dependence of $\sigma_0\crit$ on $\Delta_0$
for both uniformly and Weibull distributed strength thresholds.

In Fig.~\ref{fig:pnc_pb_bimodal_weibull_sigma0} and
\ref{fig:pb_bimodal_weibull_delta0}, our analytical results are
compared with those obtained from Monte-Carlo simulations. To ensure the validity of
condition~(\ref{eq:redist_req}), we have chosen to keep $\Delta_0$
fixed as the number~$N_\mathrm{in}$ of intact fibers decreases, and to
use the scaling $p_0 = \Delta_0/N_\mathrm{in}$. 

To compare the results of this model ($\Delta_0$-model) with those of
the model analyzed in Section 4 ($\gamma$-model), we first make the
following observations. The $\gamma$-model reproduces the GLS-limit if
$\gamma \le 2$ and (in a particular sense) approaches an LLS-limit as
$\gamma\to\infty$, while in the $\Delta_0$-model, the GLS-limit is
reproduced if $\Delta_0\to0$ and the LLS-limit (transfer of the
failing load to a single surviving fiber) for $\Delta_0\to1$. Because
of the different nature of the two $\Delta$-distributions, an exact
relation between $\gamma$ and $\Delta_0$ cannot be derived. A
comparison of Figs.~\ref{fig:sigma0crit_power_law} and
\ref{fig:sigma0crit_bimodal}, however, suggests that for $2 \le \gamma
\le 6$, a rough correspondence between the $\gamma$- and the
$\Delta_0$-model is obtained if we set
\begin{equation}
  \label{eq:gamma_vs_delta0}
  2/\gamma \approx 1 - \Delta_0\,.
\end{equation}
Using this relation, it can be seen that the $\Prob\nc$
vs. $\sigma_0$ and $\Prob\b$ vs. $\sigma_0$ behavior of the two models
is qualitatively very similar if $2 \le \gamma \le 6$ ($0 \le \Delta_0
\le 0.7$), except that in the $\gamma$-model, $\Prob\b(\sigma_0$) always increases towards
one, while in the $\Delta_0$-model, $\Prob\b(\sigma_0)$ saturates at a value smaller than
one if $\Delta_0 > 0.1$.

For $\gamma > 6$ ($\Delta_0 > 0.7$), however, the behavior of the two
models is significantly different. The critical strength
$\sigma_0\crit$, e.g., continues to decrease towards zero in the
$\Delta_0$-model, but starts to increase again in the
$\gamma$-model. Also the behavior of the no-cascade probability
$\Prob\nc$ vs. $\sigma_0$ is completely different in the two
models. In the $\gamma$-model, $\Prob\nc$ continues to increase
towards one as $\gamma \to \infty$, even for large values of
$\sigma_0$, while in the $\Delta_0$-model, $\Prob\nc$ is bounded by
$\Prob\nc < 1/\mathrm{e}$ for large $\sigma_0$. As already discussed
in Sect.~\ref{sec:results_power_law_uniform}, the peculiar behavior of
the $\gamma$-model for large $\gamma$ can be attributed to a breakdown
of the continuum approximation on which the corresponding
$\Delta$-distribution is based.

\section{Conclusions}
\label{sec:conclusions}

In this paper, we have introduced and analyzed a new fiber-bundle
model with stochastic load redistribution. The fraction~$\Delta$ of a
failing fiber that is transferred to the surviving fibers is assumed
to be a random variable, and we have considered two different
distributions for the $\Delta$-values. The first ($\gamma$-model)
refers to a stochastic version of the range-dependent load
redistribution model of Hidalgo et al.~\cite{Hidalgo2002}, and the
second ($\Delta_0$-model) to a model with a simple bimodal
$\Delta$-distribution that can also interpolate between the two
limiting cases of global and local load sharing. For the
distribution of strength thresholds, we have also considered two
different cases, a uniform and a Weibull distribution, both truncated
below some finite stress~$\sigma_0$.

While our models neglect any spatial correlations in the load
redistribution after a failure, they have the advantage that they can
be treated analytically, in contrast to most of the existing fiber
bundle models that can only be analyzed via Monte-Carlo simulations.

In the limit of global load sharing ($\gamma<2$ in the $\gamma$-model
or $\Delta_0 \to 0$ in the $\Delta_0$-model), our models not only
recover the known exact results for the critical stress~$\sigma_0\crit$, but also give
the exact behavior of the breakdown probability for $\sigma_0 >
\sigma_0\crit$. In this GLS limit, the recursion relations for the
determination of $\Prob\b(\sigma_0)$ are reduced to simple transcendental
equations
\begin{equation}
  \label{eq:pb_sigam0_uniform}
  \Prob\b(\sigma_0) = 1 - \exp\bigg[-\frac{\sigma_0}{1-\sigma_0}  \, \Prob\b(\sigma_0)\bigg]
\end{equation}
for a truncated uniform strength-threshold distribution, and
\begin{equation}
  \label{eq:pb_sigam0_weibull}
  \Prob\b(\sigma_0) = 1 - \exp\big[-2 \sigma_0^2  \, \Prob\b(\sigma_0)\big]
\end{equation}
for a truncated Weibull distribution with index $k =
2$. Eqs.~\eqref{eq:pb_sigam0_uniform} and \eqref{eq:pb_sigam0_weibull}
can easily be derived from the recursion of
Eq.~\eqref{eq:recursion_pb_bimodal} by taking the
limit~$\Delta_0\to0$, or from the transcendental
equation~(\ref{eq:pb_iter_gamma3a}) for $\gamma\le2$.

With our stochastic models, we can also determine the critical
stress~$\sigma_0\crit$ and the behavior of $\Prob\b(\sigma_0$) for
$\sigma_0 > \sigma_0\crit$ in the case of a more localized stress
redistribution ($\gamma>2$ in the $\gamma$-model or $\Delta_0 > 0$ in
the $\Delta_0$-model). As already discussed in
Sect.~\ref{sec:results_power_law_uniform}, our $\gamma$-model results
for $\sigma_0\crit$ (for truncated uniform strength-threshold
distributions) agree very well with the corresponding results of
Raischel et al.~\cite{Raischel2006} up to $\gamma \approx 6$ or
$7$. It is quite remarkable that a stochastic model that neglects any
spatial correlations can so accurately reproduce the behavior of a
microscopically more adequate model. In addition, our analytical
solution allows us to trace back the onset of the transition between
the GLS and LLS behavior at $\gamma=2$ to a change in the scaling of
the upper cutoff~$\Delta_\mathrm{max}$ of the $\Delta$-distribution in
the limit of infinite system sizes.

In the case of strength-threshold distributions that are not
truncated, the usual procedure is to gradually increase the external
force~$F$ from zero up to the complete breakdown of the entire
bundle. The critical strength of the fiber bundle is then defined as
the maximum stress~$F/N$ the system can support before it breaks down.

In global load sharing models, the surviving fibers always carry the
same stress, so that the critical fiber-bundle strength can be written
as
\begin{equation}
  \label{eq:sigma_crit_bundle}
  \sigmacritbundle = \sigma_\mathrm{c}\, [ 1 - F\c(\sigma_\mathrm{c})]\,,
\end{equation}
where $\sigma_\mathrm{c}$ is the stress a surviving fiber carries at
breakdown and $1-F\c(\sigma_\mathrm{c})$ is the fraction of surviving
fibers. This result is also recovered within our approach, with
$\sigma_\mathrm{c} = \sigma_0\crit$.

For nearest-neighbor LLS models, however, $\sigmacritbundle$ vanishes in the limit of large system
sizes, $\sigmacritbundle\propto 1/\sqrt{N}$. As our stochastic models
neglect spatial correlations, they cannot describe such situations.

We can, however, compare the results of our models with that of
Ref.~\cite{Dalton2009}, where also a stochastic load redistribution
model is used. Here, $\sigmacritbundle$ remains finite, even if the
failing load is transferred only to a small, fixed number
($n=1,2,\dots$) of randomly chosen surviving fibers. For $n=2$ and for
a uniform strength-threshold distribution, e.g., it is found that
$\sigmacritbundle\approx 0.2$ for large systems. This can be compared
with the corresponding result of our $\Delta_0$-model. If we choose
$\Delta_0=0.5$, so that the failing load is, on average, transferred
to two surviving fibers, we obtain $\sigma_0\crit =
(1-\Delta_0)/(2-\Delta_0) = 1/3$ for a uniform distribution of
strength thresholds, i.e.,
\begin{equation}
  \label{eq:sigma_crit_bundle_un}
  \sigmacritbundle = \sigma_0\crit (1 - \sigma_0\crit) = 2/9 \approx 0.22\,.
\end{equation}
The discrepancy between the two models can be attributed to our
assumption that at breakdown all surviving fibers carry the same
stress~$\sigma_0\crit$, whereas it is shown in Ref.~\cite{Dalton2009}
that at breakdown, the surviving fibers have a broad distribution
of stresses, with a pronounced exponential tail.

It remains to be investigated whether our approach can be adapted to
correctly analyze the behavior of fiber-bundle models with a
strength-threshold distribution that is not truncated.

\end{document}